\documentclass[longauth]{aa}

\usepackage[dvips]{graphicx}
\usepackage{amsmath}
\usepackage{amssymb}
\usepackage{verbatim}
\usepackage{array}
\usepackage{txfonts}
\usepackage{natbib}
\usepackage[switch,pagewise]{lineno}

\bibpunct{(}{)}{;}{a}{}{,}

\newcommand{\celltspace}{\rule{0pt}{2.8ex}}
\newcommand{\cellbspace}{\rule[-1.4ex]{0pt}{0pt}}

\def\deg{\ensuremath{^\circ}}

\begin{document}

\title{Detection of the Small Magellanic Cloud\\ in gamma-rays with Fermi/LAT}
\titlerunning{Detection of the Small Magellanic Cloud in gamma-rays with Fermi/LAT}
\author{
A.~A.~Abdo$^{1,2}$ \and 
M.~Ackermann$^{3}$ \and 
M.~Ajello$^{3}$ \and 
L.~Baldini$^{4}$ \and 
J.~Ballet$^{5}$ \and 
G.~Barbiellini$^{6,7}$ \and 
D.~Bastieri$^{8,9}$ \and 
K.~Bechtol$^{3}$ \and 
R.~Bellazzini$^{4}$ \and 
B.~Berenji$^{3}$
% ...
}
\authorrunning{LAT collaboration}
\institute{
\inst{1}~Space Science Division, Naval Research Laboratory, Washington, DC 20375, USA\\ 
\inst{2}~National Research Council Research Associate, National Academy of Sciences, Washington, DC 20001, USA\\ 
\inst{3}~W. W. Hansen Experimental Physics Laboratory, Kavli Institute for Particle Astrophysics and Cosmology, Department of Physics and SLAC National Accelerator Laboratory, Stanford University, Stanford, CA 94305, USA\\ 
\inst{4}~Istituto Nazionale di Fisica Nucleare, Sezione di Pisa, I-56127 Pisa, Italy\\ 
\inst{5}~Laboratoire AIM, CEA-IRFU/CNRS/Universit\'e Paris Diderot, Service d'Astrophysique, CEA Saclay, 91191 Gif sur Yvette, France\\ 
\inst{6}~Istituto Nazionale di Fisica Nucleare, Sezione di Trieste, I-34127 Trieste, Italy\\ 
\inst{7}~Dipartimento di Fisica, Universit\`a di Trieste, I-34127 Trieste, Italy\\ 
\inst{8}~Istituto Nazionale di Fisica Nucleare, Sezione di Padova, I-35131 Padova, Italy\\ 
\inst{9}~Dipartimento di Fisica ``G. Galilei", Universit\`a di Padova, I-35131 Padova, Italy\\ 
\inst{10}~Istituto Nazionale di Fisica Nucleare, Sezione di Perugia, I-06123 Perugia, Italy\\ 
\inst{11}~Dipartimento di Fisica, Universit\`a degli Studi di Perugia, I-06123 Perugia, Italy\\ 
\inst{12}~Centre d'\'Etude Spatiale des Rayonnements, CNRS/UPS, BP 44346, F-30128 Toulouse Cedex 4, France\\ 
\email{knodlseder@cesr.fr} \\
\email{jean@cesr.fr}\\
\inst{13}~Department of Physics, Center for Cosmology and Astro-Particle Physics, The Ohio State University, Columbus, OH 43210, USA\\ 
\inst{14}~Dipartimento di Fisica ``M. Merlin" dell'Universit\`a e del Politecnico di Bari, I-70126 Bari, Italy\\ 
\inst{15}~Istituto Nazionale di Fisica Nucleare, Sezione di Bari, 70126 Bari, Italy\\ 
\inst{16}~Laboratoire Leprince-Ringuet, \'Ecole polytechnique, CNRS/IN2P3, Palaiseau, France\\ 
\inst{17}~Institut de Ciencies de l'Espai (IEEC-CSIC), Campus UAB, 08193 Barcelona, Spain\\ 
\inst{18}~INAF-Istituto di Astrofisica Spaziale e Fisica Cosmica, I-20133 Milano, Italy\\ 
\inst{19}~NASA Goddard Space Flight Center, Greenbelt, MD 20771, USA\\ 
\inst{20}~Center for Research and Exploration in Space Science and Technology (CRESST) and NASA Goddard Space Flight Center, Greenbelt, MD 20771, USA\\ 
\inst{21}~Department of Physics and Center for Space Sciences and Technology, University of Maryland Baltimore County, Baltimore, MD 21250, USA\\ 
\inst{22}~George Mason University, Fairfax, VA 22030, USA\\ 
\inst{23}~Laboratoire de Physique Th\'eorique et Astroparticules, Universit\'e Montpellier 2, CNRS/IN2P3, Montpellier, France\\ 
\inst{24}~Department of Physics, Stockholm University, AlbaNova, SE-106 91 Stockholm, Sweden\\ 
\inst{25}~The Oskar Klein Centre for Cosmoparticle Physics, AlbaNova, SE-106 91 Stockholm, Sweden\\ 
\inst{26}~Royal Swedish Academy of Sciences Research Fellow, funded by a grant from the K. A. Wallenberg Foundation\\ 
\inst{27}~CNRS/IN2P3, Centre d'\'Etudes Nucl\'eaires Bordeaux Gradignan, UMR 5797, Gradignan, 33175, France\\ 
\inst{28}~Universit\'e de Bordeaux, Centre d'\'Etudes Nucl\'eaires Bordeaux Gradignan, UMR 5797, Gradignan, 33175, France\\ 
\inst{29}~Department of Physical Sciences, Hiroshima University, Higashi-Hiroshima, Hiroshima 739-8526, Japan\\ 
\inst{30}~Agenzia Spaziale Italiana (ASI) Science Data Center, I-00044 Frascati (Roma), Italy\\ 
\inst{31}~INAF Istituto di Radioastronomia, 40129 Bologna, Italy\\ 
\inst{32}~Center for Space Plasma and Aeronomic Research (CSPAR), University of Alabama in Huntsville, Huntsville, AL 35899, USA\\ 
\inst{33}~Instituci\'o Catalana de Recerca i Estudis Avan\c{c}ats (ICREA), Barcelona, Spain\\ 
\inst{34}~Research Institute for Science and Engineering, Waseda University, 3-4-1, Okubo, Shinjuku, Tokyo, 169-8555 Japan\\ 
\inst{35}~Department of Physics, University of Washington, Seattle, WA 98195-1560, USA\\ 
\inst{36}~Max-Planck Institut f\"ur extraterrestrische Physik, 85748 Garching, Germany\\
\email{martinp@mpe.mpg.de} \\
\inst{37}~Department of Physics and Department of Astronomy, University of Maryland, College Park, MD 20742, USA\\ 
\inst{38}~Istituto Nazionale di Fisica Nucleare, Sezione di Roma ``Tor Vergata", I-00133 Roma, Italy\\ 
\inst{39}~Department of Physics and Astronomy, University of Denver, Denver, CO 80208, USA\\ 
\inst{40}~Hiroshima Astrophysical Science Center, Hiroshima University, Higashi-Hiroshima, Hiroshima 739-8526, Japan\\ 
\inst{41}~Institute of Space and Astronautical Science, JAXA, 3-1-1 Yoshinodai, Sagamihara, Kanagawa 229-8510, Japan\\ 
\inst{42}~Institut f\"ur Astro- und Teilchenphysik and Institut f\"ur Theoretische Physik, Leopold-Franzens-Universit\"at Innsbruck, A-6020 Innsbruck, Austria\\ 
\inst{43}~Santa Cruz Institute for Particle Physics, Department of Physics and Department of Astronomy and Astrophysics, University of California at Santa Cruz, Santa Cruz, CA 95064, USA\\ 
\inst{44}~Space Sciences Division, NASA Ames Research Center, Moffett Field, CA 94035-1000, USA\\ 
\inst{45}~NYCB Real-Time Computing Inc., Lattingtown, NY 11560-1025, USA\\ 
\inst{46}~Department of Chemistry and Physics, Purdue University Calumet, Hammond, IN 46323-2094, USA\\ 
\inst{47}~Partially supported by the International Doctorate on Astroparticle Physics (IDAPP) program\\ 
\inst{48}~Consorzio Interuniversitario per la Fisica Spaziale (CIFS), I-10133 Torino, Italy\\ 
\inst{49}~INTEGRAL Science Data Centre, CH-1290 Versoix, Switzerland\\ 
\inst{50}~Dipartimento di Fisica, Universit\`a di Roma ``Tor Vergata", I-00133 Roma, Italy\\ 
\inst{51}~Department of Physics, Royal Institute of Technology (KTH), AlbaNova, SE-106 91 Stockholm, Sweden\\ 
\inst{52}~School of Pure and Applied Natural Sciences, University of Kalmar, SE-391 82 Kalmar, Sweden\\ 
}
\date{Received: 23 April 2010 / Accepted: 01 August 2010}
\abstract{The flux of gamma rays with energies greater than 100\,MeV is dominated by diffuse emission coming from cosmic-rays (CRs) illuminating the interstellar medium (ISM) of our Galaxy through the processes of Bremsstrahlung, pion production and decay, and inverse-Compton scattering. The study of this diffuse emission provides insight into the origin and transport of cosmic rays.}{We searched for gamma-ray emission from the Small Magellanic Cloud (SMC) in order to derive constraints on the cosmic-ray population and transport in an external system with properties different from the Milky Way.}{We analysed the first 17 months of continuous all-sky observations by the Large Area Telescope (LAT) of the Fermi mission to determine the spatial distribution, flux and spectrum of the gamma-ray emission from the SMC. We also used past radio synchrotron observations of the SMC to study the population of CR electrons specifically.}{We obtained the first detection of the SMC in high-energy gamma rays, with an integrated $>$100\,MeV flux of (3.7 $\pm$0.7) $\times$ 10$^{-8}$\,ph\,cm$^{-2}$\,s$^{-1}$, with additional systematic uncertainty of $\leq16$\%. The emission is steady and from an extended source $\sim3\deg$ in size. It is not clearly correlated with the distribution of massive stars or neutral gas, nor with known pulsars or supernova remnants, but a certain correlation with supergiant shells is observed.}{The observed flux implies an upper limit on the average CR nuclei density in the SMC of $\sim$15\% of the value measured locally in the Milky Way. The population of high-energy pulsars of the SMC may account for a substantial fraction of the gamma-ray flux, which would make the inferred CR nuclei density even lower. The average density of CR electrons derived from radio synchrotron observations is consistent with the same reduction factor but the uncertainties are large. From our current knowledge of the SMC, such a low CR density does not seem to be due to a lower rate of CR injection and rather indicates a smaller CR confinement volume characteristic size.}
\maketitle

\section{Introduction}
\label{intro}


\indent Gamma rays with energies greater than 100\,MeV are the observable manifestations of various populations of non-thermal particles. The sky in this energy range was unveiled by \mbox{SAS-2} and \mbox{COS-B} in the 1970s, then mapped by CGRO/EGRET in the 1990s, and is now surveyed by AGILE since 2007 and {\em Fermi}/LAT since 2008. At low Galactic latitudes, many supernova remnants (SNRs) and pulsars have revealed themselves as accelerators of particles up to TeV energies, but a lot of localised sources are still formally unidentified. Most of the emission, however, is of diffuse nature and arises from the interaction of Galactic cosmic-rays (CRs) with interstellar matter and radiation. The pervasiveness of this emission, which accounts for 80-90\% of the flux above 100\,MeV, indicates that CRs diffuse far from their sources filling the entire Galaxy.\\
\indent Since their discovery at the beginning of the 20th century, it has been realised that CRs are not simply a side-product of the most energetic phenomena of the Universe, but actually form an essential component of the Milky Way (MW). Galactic CRs affect the physical state of the interstellar medium (ISM) by heating/ionising its atomic and molecular phases \citep{Ferriere:2001} and are thought to alter the spectrum of interstellar turbulence as they get reaccelerated by draining energy from magnetohydrodynamic waves \citep{Ptuskin:2006}. Ionisation by low-energy CRs is important for the chemical processes operating inside dense molecular clouds, where Lyman photons cannot penetrate \citep{Indriolo:2009}. Because of their high energy density \mbox{($\sim$ 1\,eV\,cm$^{-3}$)} and corresponding pressure, CRs also contribute to supporting interstellar gas against gravity \citep{Ferriere:1998}. In addition, CRs are essential for the Parker dynamo mechanism, which may be the source of the large-scale magnetic fields in galaxies like the Milky Way \citep{Hanasz:2004,Hanasz:2009}.\\
\indent Direct measurements of the CR composition and spectrum in the interplanetary medium, together with imaging of the Galactic diffuse emission in the radio and MeV-GeV domain, have provided a wealth of information on Galactic CRs. Combined with models of increasing sophistication, from simple leaky box to analytic diffusion models \citep[e.g.][]{Aharonian:2000} to more elaborate numerical calculations \citep[such as GALPROP, see][]{Strong:2004}, these data permitted deriving important constraints on the transport of CRs inside our Galaxy: the time spent in the Galactic volume, the amount of matter traversed, the extent of spatial diffusion, and the role of reacceleration and convection \citep[see the review by][]{Strong:2007}.\\
\indent Resolved gamma-ray observations of externally-viewed galaxies provide the possibility to make further progress understanding the origin and propagation of CRs. The Magellanic Clouds offer a great opportunity in this regard due their proximity and their significantly contrasting geometries and physical conditions (gas, stars, magnetic field, ...) compared to the Milky Way. The Large Magellanic Cloud (LMC) was observed by EGRET with a measured flux $>$100\,MeV of (1.9 $\pm$0.4) $\times$ 10$^{-7}$\,ph\,cm$^{-2}$\,s$^{-1}$ that was found to be consistent with a model in which CRs interact with the ISM of the galaxy, under the assumption that the CRs, magnetic pressure, and thermal gas are in dynamical balance with the gravitational attraction \citep{Fichtel:1991,Sreekumar:1992}. While it was clear that the LMC was an extended object, the sensitivity of EGRET precluded resolving any features. The LMC has now been resolved by {\em Fermi}/LAT. The higher sensitivity and angular resolution of the instrument revealed a good correlation between gamma-ray emission and massive star forming regions, thereby providing evidence for CR acceleration in these regions and suggesting a relatively small diffusion length for GeV CRs \citep{Abdo:2009g}. In contrast, the Small Magellanic Cloud (SMC) was not detected by EGRET but the upper-limit thus derived allowed to dismiss a metagalactic origin for the bulk of CRs with energies 10$^{15}$-10$^{16}$\,eV \citep{Sreekumar:1993}, while the most energetic particles are thought to come from extragalactic sources lying within the Greisen-Zatsepin-Kuz'min (GZK) horizon \citep{PierreAugerCollaboration:2007,PierreAugerCollaboration:2008,PierreAugerCollaboration:2009}.\\
\indent In this paper, we report the detection and resolving of the SMC in high-energy gamma-rays using {\em Fermi}/LAT. In the following, we describe the various analyses performed on the {\em Fermi}/LAT data in order to derive the morphology, the flux and the spectrum of the SMC in gamma-rays. We then discuss plausible origins for the detected emission before considering the consequences of our result on the CR population and transport in the SMC.

\section{Observations}
\label{obs}

\subsection{Data preparation}
\label{obs_data}

\indent The characteristics and performance of the LAT aboard {\em Fermi} are described in detail by \citet{Atwood:2009}. We used the LAT Science Tools package, which is available from the Fermi Science Support Center, with the P6\_V3 post-launch instrument response functions (IRFs). These take into account pile-up and accidental coincidence effects in the detector subsystems that were not considered in the definition of the pre-launch IRFs.\\
\indent For the analysis, we selected all events within a square region-of-interest (ROI) of size $20\deg \times 20\deg$ centred on $(l,b)=(302.8\deg,-44.3\deg)$ and aligned in Galactic coordinates\footnote{All the maps presented in this article are in Galactic coordinates. To ease the comparison with other studies of the SMC, the same maps are provided in equatorial coordinates as online material.}. The data used in this work amount to 504 days of sky survey observations over the period August 8th 2008 -- December 25th 2009 during which a total exposure of $\sim4.5 \times 10^{10}$~cm$^2$~s (at 1 GeV) was obtained for the SMC. Events of the \textit{Diffuse} class \citep[which are the least contaminated by the cosmic-ray background, see][]{Atwood:2009} and coming from zenith angles $<105\deg$ \citep[to greatly reduce the contribution by Earth albedo gamma rays, see][]{Abdo:2009m} were used. To further reduce the effect of Earth albedo, the time intervals when the Earth was appreciably within the field of view (specifically, when the centre of the field of view was more than $52\deg$ from the zenith) were excluded from this analysis. We further restricted the analysis to photon energies above 200\,MeV because the effective area for \textit{Diffuse} class events changes rapidly at lower energies, and up to 20\,GeV because the number of counts is quite low beyond this limit.

\subsection{Background modelling}
\label{obs_back}

\indent At the mid-latitude position of the SMC ($b \sim -45\deg$), the gamma-ray background is a combination of extragalactic and Galactic diffuse emissions. The extragalactic component comprises resolved sources, which often can be associated with known blazars and other extragalactic source classes, and a diffuse component, which is attributed to unresolved sources and eventually intrinsically diffuse processes \citep{Abdo:2009c,Abdo:2010a}.\\
\indent The 1FGL catalog \citep{Abdo:2010b} has been derived from 11 months of survey data (comparable to the data volume used in this analysis) and contains 622 sources for latitudes $|b| \ge 30\deg$, corresponding to a source density of $99$ sources per steradian. Our ROI covers a solid angle of 0.12 sr, thus we expect about 12 resolved background sources in our field. Among those, $1-2$ should spatially overlap with the SMC for an assumed angular size $\sim6-8\deg$ for the galaxy.\\
\begin{figure}[!t]
\begin{center}
\includegraphics[width= \columnwidth]{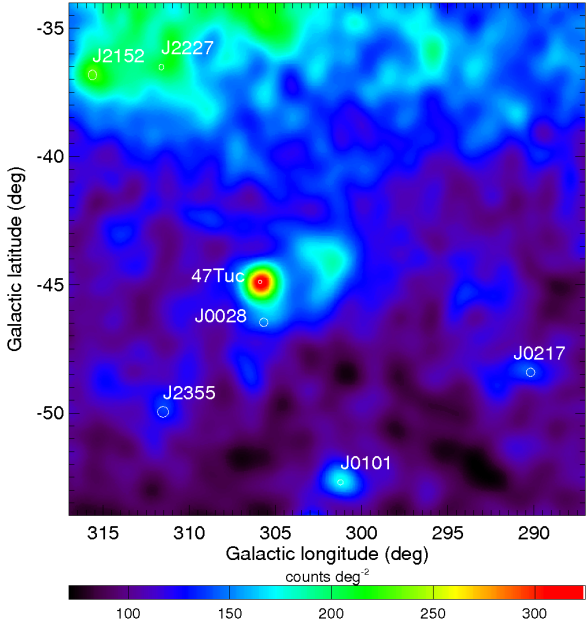}
\caption{200\,MeV$-$20\,GeV counts map of the region of interest centred on the SMC, after smoothing with a 2D Gaussian kernel with $\sigma = 0.4\deg$. The positions of the background point sources are marked by white circles, with circle size indicating position uncertainty}
\label{fig_cntmap_roi}
\end{center}
\end{figure}
\begin{figure}[!t]
\begin{center}
\includegraphics[width= \columnwidth]{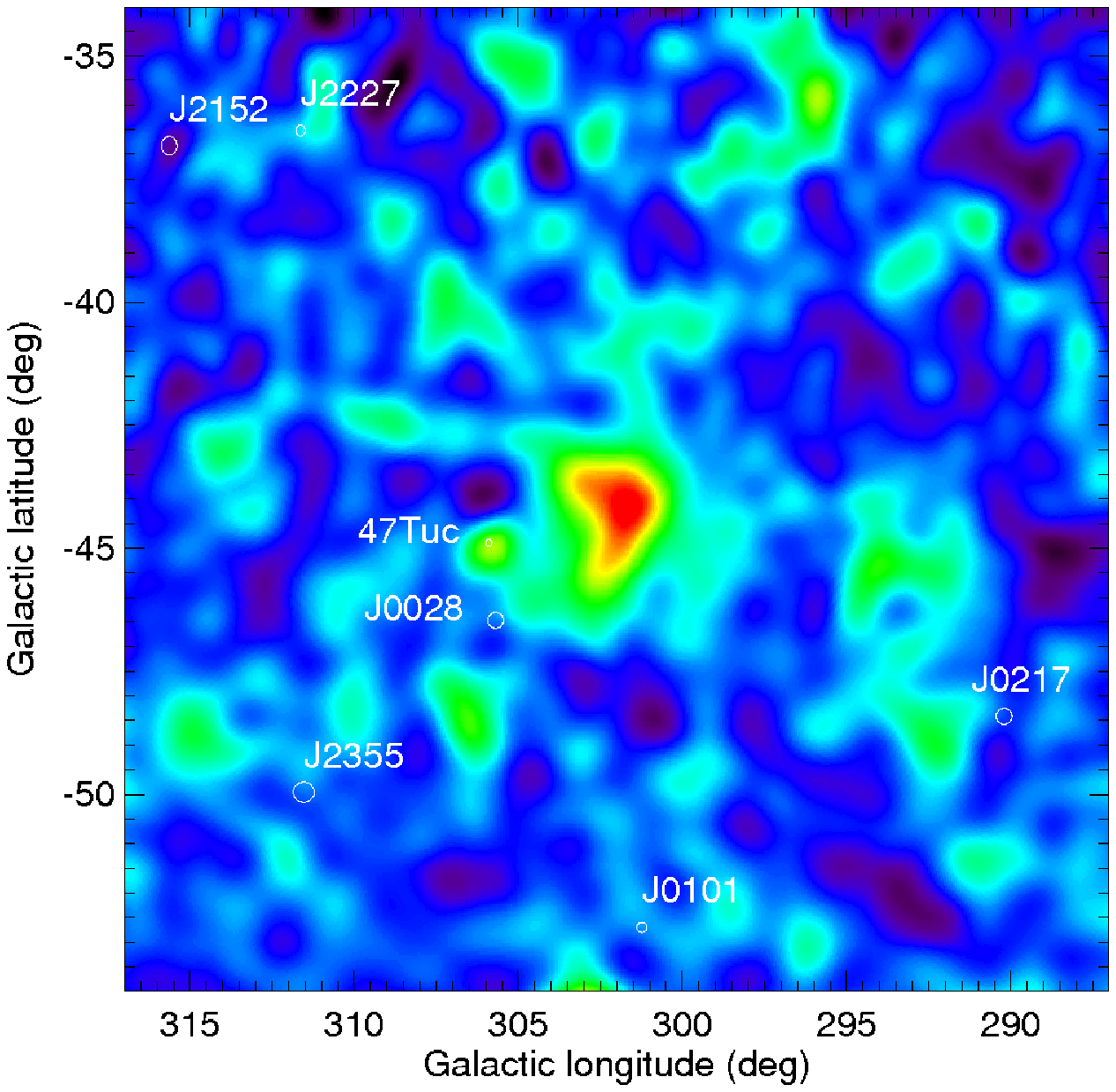}
\includegraphics[width= \columnwidth]{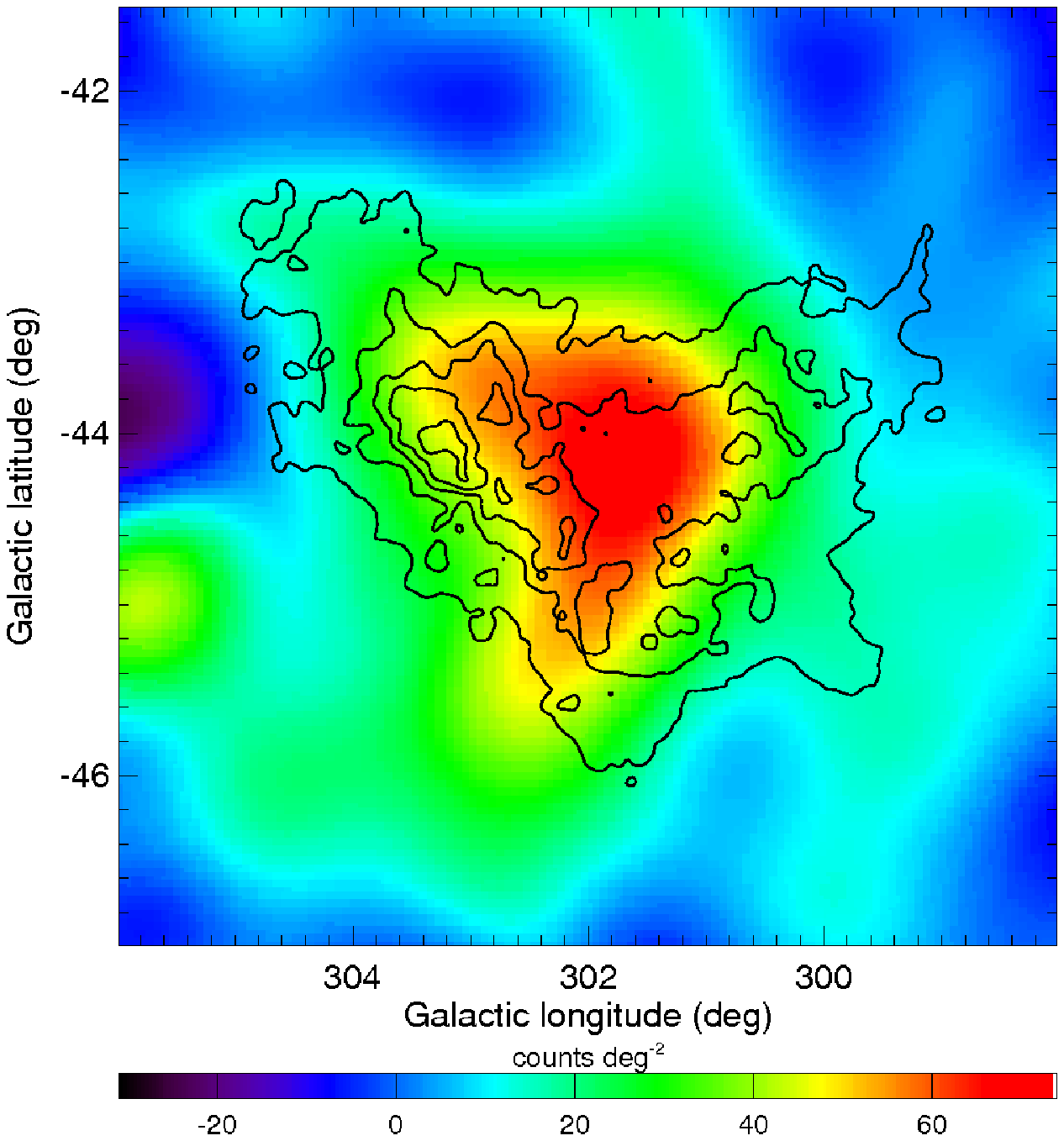}
\caption{200\,MeV$-$20\,GeV residual counts maps after subtraction of the fitted celestial background model and a smoothing with a 2D Gaussian kernel with $\sigma = 0.4\deg$. Upper plot corresponds to the full region of interest and lower plot is a zoom on the SMC. The positions of the background point sources are marked by white circles in the top panel, with circle size indicating position uncertainty, and the SMC is traced by H\,{\scriptsize I} column density contours in the bottom panel.}
\label{fig_resmap}
\end{center}
\end{figure}
\indent Within the ROI but outside the boundaries of the SMC, we found a total of 7 significant point sources, in rough agreement with expectations. One of these is the globular cluster 47Tuc, which was found to be a strong source of gamma-rays, presumably due to its population of millisecond pulsars \citep{Abdo:2009f}. Although we found no counterparts for the other 1FGL sources in the CRATES catalogue of flat-spectrum radio sources \citep{Healey:2007} using the procedure described in \citet{Abdo:2009b}, they are likely to be background blazars due to their high latitude positions. Searching for background sources within the SMC boundaries is more difficult due to possible confusion with point sources in the galaxy itself or with local emission maxima in its diffuse emission. However, background blazars may reveal themselves by their flaring activity. We therefore searched for excess emission above the time-averaged level on a monthly basis but found no indication for a flare coming from the direction of the SMC (see Sect. \ref{emiss_var}).\\
\indent We modelled the celestial background emission within the ROI by fitting components for the diffuse Galactic and extragalactic backgrounds, 47Tuc, and the 6 background objects from the 1FGL catalogue.\\
\indent The LAT standard background model {\tt gll\_iem\_v02} for the Galactic diffuse emission was found to be inadequate for the specific SMC region. Subtracting this fitted model (and the other background components described below) from the data left strong positive residuals at low latitudes ($b \sim -35\deg$) that were found to be correlated with other tracers of the interstellar gas, such as the E(B$-$V) reddening map built by \citet{Schlegel:1998} from COBE/DIRBE and IRAS/ISSA submillimeter observations of dust emission \citep[the suitability of E(B$-$V) as gas tracer was illustrated for instance in][]{Abdo:2009h}. A dedicated model for the Galactic diffuse emission in the SMC region, based on this E(B$-$V) map and an inverse-Compton (IC) intensity map computed using the GALPROP code for CR propagation in the Galaxy \citep{Strong:2000} and its model for the interstellar radiation field \citep{Porter:2008}, turned out to improve the situation and yielded a better likelihood. The E(B$-$V) map is converted to hydrogen column density map using the best-fit relation obtained by \citet{Grenier:2005}, and a $5\deg \times 3\deg$ region around the SMC is excluded and filled by the inpainting method described by \citet{Elad:2005}. As such, it is intended to trace the emission due to CRs interacting with interstellar matter in our Galaxy, and it is fitted assuming a power-law spectral shape with normalisation and spectral index as free parameters. The IC map\footnote{From GALPROP run {\tt 54\_87Xexph7S}, which is available from "http://galprop.stanford.edu".} gives the emission arising from CRs interacting with interstellar radiation and is fitted with the spectral shape as calculated with GALPROP.\\
\indent An isotropic background component was fitted assuming a power-law spectral shape with normalisation and spectral index as free parameters. This component is designed to account for the diffuse/unresolved extragalactic emission, along with residual instrumental background.\\
\indent The 6 background 1FGL objects are modelled as point sources at the positions given in the catalog and having power-law spectral shapes with normalisation and spectral indices as free parameters. Last, the globular cluster 47Tuc lying close to the SMC (in the plane of the sky) was modelled as a point source at its SIMBAD position with an exponentially cutoff power-law spectral shape with normalisation, spectral index, and cutoff energy as free parameters \citep[in agreement with the dedicated study of][]{Abdo:2009f}. The characteristics of our complete celestial background model are listed in Table \ref{tab_background}.

\section{Emission morphology}
\label{morpho}

\indent To investigate the spatial distribution of gamma-ray emission toward the SMC, we first binned all photons into a counts map of size $20\deg \times 20\deg$ centred on $(l,b)=(302.8\deg,-44.3\deg)$ and aligned in Galactic coordinates. Figure \ref{fig_cntmap_roi} shows the counts map for the full ROI, with a pixel size of $0.05\deg \times 0.05\deg$ and a smoothing of the statistical fluctuations with a 2D Gaussian kernel of width $\sigma=0.4\deg$. The region is clearly dominated by 47Tuc close to the centre, and by the Galactic diffuse emission at low latitudes.\\
\indent The top panel of Fig. \ref{fig_resmap} shows the counts map for the full ROI after subtraction of the background model (by a binned maximum likelihood analysis with 10 logarithmically-spaced energy bins), with a pixel size of $0.05\deg \times 0.05\deg$ and a smoothing of the statistical fluctuations with a 2D Gaussian kernel of width $\sigma=0.4\deg$. The map suggests that the background is properly removed by our treatment. In particular, the intensity gradient due to the diffuse Galactic emission increasing as we move toward the Galactic plane as well as the 6 background 1FGL objects and 47Tuc have disappeared. The only remaining feature is a clear excess at the position of the SMC. We emphasise here that, whatever the Galactic diffuse model used (the standard {\tt gll\_iem\_v02} model or our dedicated one), residual emission at the position of the SMC was observed and the excess counts were distributed in a very similar way. In addition, the characteristics of this residual emission were found to be insensitive to the inpainting or interpolation method used in the preparation of the E(B$-$V) map for the dedicated model.\\
\indent The bottom panel of Fig. \ref{fig_resmap} is a zoom on the centre of the ROI showing residual emission over a few degrees and spatially coincident with the SMC, as traced by H\,{\scriptsize I} column density contours \citep[from][]{Stanimirovic:1999}. In Sect. \ref{morpho_geom}, we show that the extension of this excess emission is real, and not just an effect of the IRF. The total number of excess 200 MeV -- 20 GeV photons above the background in a $4\deg \times 4\deg$ square covering the residual excess amounts to $\sim 450$ counts, whereas the background in the same area amounts to $\sim 1820$ counts. Such modest statistics obviously limit the precision of the conclusions about the spatial distribution of the extended emission from the SMC.\\
\indent The residual counts map shows several spots with intensities of $\sim$40-50\,counts/deg$^{2}$, similar to the peak residual intensity associated with the SMC ($\sim$65\,counts/deg$^{2}$). None of these spots actually is significant except for the excess at $(l,b) \sim (296\deg,-36\deg)$ which is at the $\sim 4 \sigma$ level. In contrast, the significance of the SMC emission is $\sim 11 \sigma$ (see Sect. \ref{morpho_geom}), mainly because it is significantly extended while the other hot spots of the residual counts map are mostly point-like. We also checked that adding these hot spots as point sources in our model fitting did not alter the characteristics derived for the SMC emission.
\begin{table*}[!t]
\begin{minipage}[][7.5cm][c]{\textwidth}
\begin{center}
\caption{Spatial and spectral characteristics of each component of the celestial background model used in the analysis.}
\begin{tabular}{|c|c|c|c|}
\hline
\celltspace Component & Spatial model & Spectral model & Parameters \cellbspace \\
\hline
\celltspace Galactic diffuse & E(B$-$V) map from \citet{Schlegel:1998} & power law & 2 \cellbspace\\
					  & IC map from GALPROP & from GALPROP & 1 \cellbspace \\
 \hline
\celltspace Extragalactic diffuse & Isotropic & power law & 2 \cellbspace \\
 \hline
\celltspace 47Tuc (1FGL J0023.9-7204) & point source at $(l,b)=(305.89\deg, -44.89\deg)$ & \hspace{0.5cm} power law + cutoff \hspace{0.5cm} & 3 \cellbspace \\
 \hline
\celltspace \hspace{0.5cm} 1FGL J0028.9-7028 \hspace{0.5cm} & \hspace{0.5cm} point source at $(l,b)=(305.66\deg, -46.53\deg)$ \hspace{0.5cm} & power law & 2 \cellbspace \\
 \hline
\celltspace 1FGL J0101.0-6423 & point source at $(l,b)=(301.22\deg, -52.70\deg)$ & power law & 2 \cellbspace \\
 \hline
\celltspace 1FGL J0217.9-6630 & point source at $(l,b)=(290.14\deg, -48.42\deg)$ & power law & 2 \cellbspace \\
 \hline
\celltspace 1FGL J2152.4-7532  & point source at $(l,b)=(315.62\deg, -36.82\deg)$ & power law & 2 \cellbspace \\
 \hline
\celltspace 1FGL J2227.4-7804  & point source at $(l,b)=(311.62\deg, -36.51\deg)$ & power law & 2 \cellbspace \\
 \hline
\celltspace 1FGL J2355.9-6613 & point source at $(l,b)=(311.57\deg, -49.96\deg)$ & power law & 2 \cellbspace \\
\hline
\end{tabular}
\label{tab_background}
\end{center}
\end{minipage}
\end{table*}

\subsection{Geometrical models}
\label{morpho_geom}

\indent We assessed the spatial distribution of the SMC emission using simple parameterized geometrical models for the gamma-ray intensity distribution. We considered point-like and 2D Gaussian models with free geometric parameters, and for each model we assumed a power-law spectrum with normalisation and spectral index also treated as free parameters. In our procedure, the spatial and spectral parameters of the models are adjusted using a binned maximum likelihood analysis with spatial pixels of $0.1\deg \times 0.1\deg$ and 10 logarithmically spaced energy bins covering the energy range 200 MeV -- 20 GeV. For each model of the SMC, we computed the so-called {\em Test Statistic} (TS) which is defined as twice the difference between the log-likelihood $\mathcal{L}_1$ that is obtained by fitting the model on top of the background model to the data, and the log-likelihood $\mathcal{L}_0$ that is obtained by fitting the background model only, i.e. ${\rm TS} = 2(\mathcal{L}_1 - \mathcal{L}_0)$. Under the hypothesis that the background model satisfactorily explains our data, and if the number of counts is high enough, the TS follows a $\chi^2_p$ distribution with $p$ degrees of freedom, where $p$ is the number of additional free parameters in the maximization of $\mathcal{L}_1$ with respect to those used in the maximisation of $\mathcal{L}_0$ (in the present case, $p$ is the number of free parameters of our model of the SMC). If, however, the TS takes a value that is statistically unlikely for a $\chi^2_p$ distribution, this means that the background model is not an adequate enough representation of the observations and can be improved by the additional SMC component \citep[see][]{Cash:1979}.\\ 
\indent We first aimed at explaining the gamma-ray emission from the SMC by a combination of point sources. For this purpose we successively added point sources to our model and optimised at each step the location, flux, and spectral index of the new component by maximisation of the likelihood criterion. We stopped the procedure when adding a new point source did not improve the TS value by more than 10 (which formally corresponds to a 2$\sigma$ detection for an additional model component with 4 degrees of freedom). From this method, we obtained a model of the residual emission from the SMC region consisting of 3 point sources and yielding a TS of 130.4 for a total number of 12 free parameters. This model will be referred to as 3PS hereafter. The strongest point source (PS1) is located at $(l,b)=(301.7\deg \pm0.2\deg, -44.4\deg \pm0.2\deg)$ and alone gives a TS of 98.2 for 4 degrees of freedom. We note here that the position of PS1 approximately corresponds to the position of the source 1FGL J0101.3-7257 that is associated with the SMC in the catalog. The second point source (PS2) is located at $(l,b)=(303.4\deg \pm0.3\deg, -43.6\deg \pm0.3\deg)$ and increases the TS to 119.2 for an additional 4 degrees of freedom. The third point source (PS3) is located at $(l,b)=(302.1\deg \pm0.6\deg, -45.4\deg \pm0.5\deg)$ and further increases the TS to 130.4 for 4 more degrees of freedom. Quoted uncertainties correspond to 2$\sigma$ intervals. These point sources correspond to the strongest maxima in the residual emission from the SMC and their positions may mark specific sites or objects in the galaxy: strong concentrations of CRs and/or gas if gamma-rays come primarily from CR-ISM interactions, or high-energy objects like pulsars. The maxima also could be statistical fluctuations.\\
\indent Next, instead of using a set of point sources we considered a 2D Gaussian shaped intensity profile to build a model that is more appropriate for extended and diffuse emission structures. Due to limited statistics, we restricted ourselves to a single Gaussian (we will see later that the photon statistics are currently insufficient to allow a more sophisticated model). The best fit 2D Gaussian has a width $\sigma= 0.9\deg \pm0.5\deg$, corresponding to a characteristic angular size of $\sim$3\deg, and is centred on $(l,b)=(302.1\deg \pm0.4\deg, -44.4\deg \pm0.4\deg)$. Quoted uncertainties correspond to 2$\sigma$ intervals. The TS obtained for this model, referred to as 2DG in the following, is 136.6, which is above the TS of the 3PS model for fewer degrees of freedom (5 instead of 12). The significance of the extension, when comparing the 2DG model with respect to the single point-source PS1 model, is above 6$\sigma$. These results suggest that the emission from the SMC is diffuse in nature.\\
\indent Finally, we tested the possibility that the gamma-ray emission from the SMC is not completely diffuse and has a point-source component, as might be expected for a pulsar adding its radiation to the diffuse emission arising for instance from CR-ISM interactions. We took this hypothetical point source to be the PS1 component of the 3PS model. Essentially, we evaluated how the inferred diffuse emission is impacted when the maximum of the residual emission from the SMC is attributed to a point-like object. In this case, the best-fit 2D Gaussian was found to be quite similar in terms of position and width, but the associated flux dropped by $\sim$20\% (see Table \ref{tab_morpho}). The TS value for this model is 138.8, a modest increase of 2.2 compared to the 2DG model. This attempt to improve our representation of the emission from the SMC actually reveals the limits on the spatial information that can be derived from the current set of observations. Yet, it also confirms that most of the emission is diffuse in nature: a higher TS is achieved for fewer degrees of freedom when complementing the PS1 component with an extended component rather than with more point sources, and in that case $\sim$80\% of the flux comes from the extended component.

\subsection{Tracer maps}
\label{morpho_maps}

\indent The morphology study presented in \ref{morpho_geom}, based on analytical intensity distributions, indicates that the emission from the SMC is of diffuse nature. This supports the idea that the observed emission can be attributed to CRs interacting with the ISM, although at this point we cannot dismiss the scenario of an ensemble of weak unresolved point sources in the galaxy like pulsars. To test the former hypothesis further, we compared our data to spatial templates that trace various components of the SMC relevant to CR-ISM interactions. In the 200\,MeV -- 20\,GeV energy range, the gamma-ray emission from CR-ISM interactions is dominated by the decay of neutral pions $\pi^0$ created by inelastic collisions between CR nuclei and interstellar gas particles \citep[see for example][]{Strong:2004}. As tracers of CR-ISM interactions, we therefore used a map of neutral atomic hydrogen, the major target for CR nuclei in the SMC (molecular hydrogen comprises $<$10\% of the gas mass in the SMC, see \ref{smcsrc_cr}), and a map of the ionised hydrogen that indirectly traces massive stars, that are putative accelerators of CRs (through their winds and/or explosions).\\
\indent For the neutral atomic hydrogen distribution, we used the map of \citet{Stanimirovic:1999} that is based on a combination of data from the ATCA and Parkes observatories. The map was converted into hydrogen column densities under the assumption of optically thin emission. The conversion of H\,{\scriptsize I} spectral line observations to H\,{\scriptsize I} column densities can actually be achieved under various hypotheses, optically thin or thick emission, with various spin temperatures in the latter case, but \citet{Stanimirovic:1999} showed that the correction for self-absorption in the SMC is a relatively small effect. In addition, the photon statistics in our gamma-ray data are very likely not high enough for the above subtleties to have an important effect on the results. The adopted scheme for the preparation of the H\,{\scriptsize I} map proved to affect only marginally the detection significance in the analysis of the {\em Fermi}/LAT observations of the diffuse gamma-ray emission of the LMC \citep[which has a 7 times greater flux than the SMC, see][]{Abdo:2009g}.\\
\indent For the ionised hydrogen tracer, we extracted from the full-sky composite H$\alpha$ map of \citet{Finkbeiner:2003} a circular region of 3\deg\ in radius around the position $(l,b)=(302.8\deg,-44.3\deg)$. A foreground Galactic contribution to the observed emission is estimated from the off-source H$\alpha$ intensity in the nearby area outside our selection, and subtracted from the circular region. We note that a more accurate tracer of massive stars in the SMC could be obtained by taking into account foreground extinction by dust in the Milky Way and internal extinction in the SMC \citep[see for instance][]{Mao:2008}, but we again emphasise that the current photon statistics of our observations does not justify such efforts.\\
\indent Both maps were convolved with the IRFs and fitted individually to the data (together with the background model) using a binned likelihood analysis with 10 energy bins and assuming a power-law spectral shape with free normalisation and spectral index for each model. This led to the following results: the H\,{\scriptsize I} map gives a TS of 120.8 while the H$\alpha$ map gives a TS of 114.4, for 2 degrees of freedom in each case. The gas distribution traces the gamma-ray emission slightly better than the massive stars distribution. We note here that, even under the assumption of gamma-ray emission from CR-ISM interactions, H\,{\scriptsize I} or H$\alpha$ distributions may not be the best representations of the emission. Indeed, an intensity distribution correlated with CR sources, such as massive stars, would mean CRs with GeV energies are confined relatively close to their sources in a uniform hydrogen density. On the other hand, an intensity distribution correlated with the CR targets would imply a uniform CR density. Yet, while the CR diffusion length may indeed be short \citep{Abdo:2009g}, we know that in the Milky Way both the gas and the CRs are not uniformly distributed and this is likely to be the case in the SMC as well.\\
\indent As in \ref{morpho_geom} for the 2D Gaussian model, we tested the possibility that the gamma-ray emission from the SMC has a composite origin and includes at least one strong point source. We again assumed this hypothetical point source to be the PS1 component of the 3PS model, and we complemented it with the tracer maps. In this case, the PS1+H\,{\scriptsize I} model gives a TS of 135.2 while the PS1+H$\alpha$ model gives a TS of 128.0. The conclusions actually are the same as in \ref{morpho_geom}: an extended component in addition to PS1 is more significant than other point sources like PS2 and PS3, and in that case $\sim$80\% of the flux is associated with that extended component (see Table \ref{tab_morpho}).
\begin{table*}[!t]
\begin{minipage}[][10.1cm][c]{\textwidth}
\begin{center}
\caption{Comparison of the different spatial models used to represent the emission from the SMC.}
\begin{tabular}{|c|c|c|c|c|c|c|c|}
\hline
\celltspace Name & Composition & Position & $\Delta$TS & Parameters & Flux & TS & Significance \cellbspace \\
\hline
\celltspace 3PS & point source 1 (PS1) & (301.7\deg, -44.4\deg) & 98.2 & 4 & 0.93 $\pm$0.26 & 130.4 & 9.6 \cellbspace \\
			 & point source 2 (PS2) & (303.4\deg, -43.6\deg) & 21.0 & 4 & 0.44 $\pm$0.22 &  & \cellbspace \\
			 & point source 3 (PS3) & (302.1\deg, -45.4\deg) & 11.2 & 4 & 0.31 $\pm$0.19 &  & \cellbspace \\
\hline
\celltspace 2DG & 2D Gaussian with $\sigma$=0.9\deg & (302.1\deg, -44.4\deg) & - & 5 & 2.11 $\pm$0.24 & 136.6 & 10.9 \cellbspace \\
\hline
\celltspace H\,{\scriptsize I} & H\,{\scriptsize I} column density map & - & - & 2 & 2.13 $\pm$0.25 & 120.8 & 10.7 \cellbspace \\
\hline
\celltspace H$\alpha$ & H$\alpha$ map & - & - & 2 & 1.94 $\pm$0.24 & 114.4 & 10.4 \cellbspace \\
\hline
\celltspace PS1+ H\,{\scriptsize I} & PS1 & (301.7\deg, -44.4\deg) & 98.2 & 4 & 0.46 $\pm$0.27 & 135.2 & 10.7 \cellbspace \\
			 	       & H\,{\scriptsize I} column density map & - & 37.0 & 2 & 1.65 $\pm$0.42 &  & \cellbspace \\
\hline
\celltspace PS1+ H$\alpha$ & PS1 & (301.7\deg, -44.4\deg) & 98.2 & 4 & 0.41 $\pm$0.18 & 128.0 & 10.4 \cellbspace \\
			 	                    & H$\alpha$ map & - & 29.8 & 2 & 1.47 $\pm$0.27 &  & \cellbspace \\
\hline
\celltspace PS1+ Gaussian & PS1 & (301.7\deg, -44.4\deg) & 98.2 & 4 & 0.39 $\pm$0.25 & 138.8 & 10.4 \cellbspace \\
			 	                & 2D Gaussian with $\sigma$=1.0\deg & (302.3\deg, -44.3\deg) & 40.6 & 5 & 1.71 $\pm$0.37 &  & \cellbspace \\
\hline
\end{tabular}
\label{tab_morpho}
\end{center}
Note to the table: From left to right, table columns are the name of the spatial model, the description of its components, the position in Galactic longitudes and latitudes of the point source and 2D Gaussian models, the TS increase due to each component, the number of degrees of freedom for each component, the flux in each component in units of 10$^{-8}$\,ph\,cm$^{-2}$\,s$^{-1}$ for photons in the 200\,MeV$-$20\,GeV range (with statistical uncertainties only), the total TS of the model, and its significance in $\sigma$. The H\,{\scriptsize I} column density map and the H$\alpha$ map are from \citet{Stanimirovic:1999} and \citet{Finkbeiner:2003}, respectively (see text).
\end{minipage}
\end{table*}

\subsection{Correlations}
\label{morpho_correl}

\begin{figure}[t]
\begin{center}
\includegraphics[width=\columnwidth]{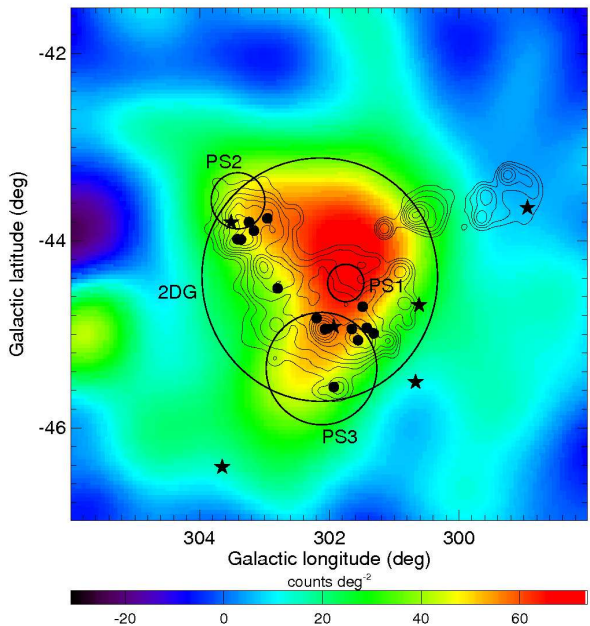}
\caption{Residual counts map after subtraction of the fitted celestial background model and smoothing with a 2D Gaussian kernel with $\sigma = 0.4\deg$. Logarithmically-spaced H$\alpha$ emission contours of the SMC are shown, together with the locations of the currently-known pulsars and SNRs in the galaxy (stars and points respectively). Black circles mark the extent of the 2DG model and the positions of the components of the 3PS model (including in all cases the 2$\sigma$ uncertainty in the position).}
\label{fig_correl}
\end{center}
\end{figure}

\indent The results obtained for the various spatial models tested to account for the SMC emission are summarised in Table \ref{tab_morpho}. The model that best fits the data is the 2DG model, which gives the highest detection significance at a level of $10.9\sigma$. Yet, the photon statistics of the set of observations used here are not sufficient to conclusively distinguish the various extended emission morphologies tested in this study.\\
\indent In Fig. \ref{fig_correl}, we plot the extent of the 2DG model, computed as the sum of the $\sigma$ of the Gaussian and the uncertainty in the position of its centroid, over the smoothed gamma-ray emission from the SMC. Although being statistically less likely than the 2DG model, the 3PS model gives indications on emission maxima that may correspond to particular sites or objects in the galaxy. We therefore marked these positions and their typical 2$\sigma$ uncertainty domains in Fig. \ref{fig_correl}. For comparison, we also show in Fig. \ref{fig_correl} the H$\alpha$ emission contours of the SMC and the locations of the currently-known pulsars and SNRs\footnote{We used the ATNF catalogue version 1.36 from the web page "http://www.atnf.csiro.au/research/pulsar/psrcat/" \citep{Manchester:2005} and SNRs data come from Rosa Williams web page "http://www.astro.illinois.edu/projects/atlas/".}.\\
\indent The position of the maximum of the emission is consistent in both the 3PS and 2DG models. Even when a hypothetical point source is placed at the emission maximum, the best-fit 2D Gaussian accounting for the remaining flux is centred at about this same position (see model PS1+Gaussian in Table \ref{tab_morpho}). From Figs. \ref{fig_resmap} and \ref{fig_correl}, this place does not seem to correspond to any particular site or object in the SMC; it is neither a strong concentration of gas or massive stars, nor the direction to one of the pulsars or SNRs of the galaxy. However, it is consistent with the centre of supergiant shell 304A \citep{Stanimirovic:1999}, located at $(l,b)=(301.8\deg,-44.5\deg)$. Although they are not highly significant (detection at the 2-3$\sigma$ level), we note that PS2 is found near the peak hydrogen density and close to the centre of supergiant shell 37A \citep{Stanimirovic:1999}, while the position of PS3 is consistent within uncertainties with NGC346, the most active star-forming region of the SMC.\\
\indent Regarding the tracers of CR-ISM interactions, both the H\,{\scriptsize I} and H$\alpha$ maps are relatively good tracers of the extended emission from the SMC. The photon statistics of the available SMC observations are not sufficient to conclusively distinguish one tracer from the other. In the case of the LMC, the gamma-ray emission is found to be strongly correlated with massive star forming regions and poorly correlated with gas density \citep{Abdo:2009g}. Additional observations of the SMC with {\em Fermi} are needed to clarify the differences and similarities in the gamma-ray emission of the two Magellanic Clouds. From these, important information will very likely be obtained about the origin and transport of the CRs.

\section{Emission lightcurve and spectrum}
\label{emiss}

\subsection{Variability}
\label{emiss_var}

\begin{figure}[t]
\begin{center}
\includegraphics[width=\columnwidth]{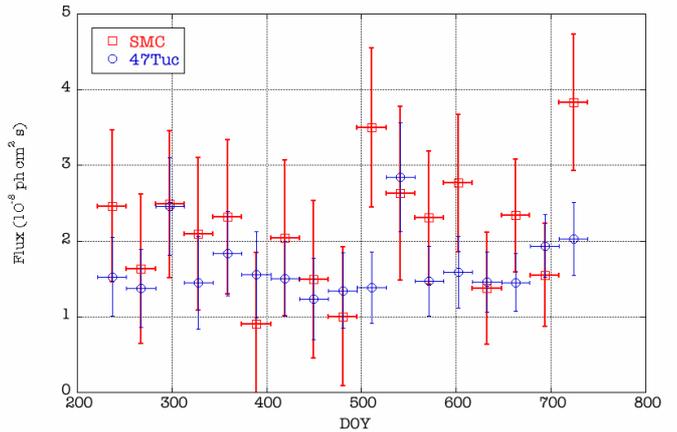}
\caption{Monthly light curve of the SMC obtained with the 2DG spatial model and a power-law spectral shape with fixed index. Also shown for comparison is the light curve of 47Tuc.}
\label{fig_light curve}
\end{center}
\end{figure}

\indent As noted earlier, emission from within the SMC boundaries could arise from background blazars and be incorrectly interpreted as local emission maxima of the diffuse emission from the galaxy. Background blazars may however reveal themselves by their flaring activity and we therefore searched for any excess emission above the time-averaged level.\\
\indent We first considered the time variability of the integrated gamma-ray emission from the SMC by using the extended 2DG model (on top of our background model). As before, the spectrum of the 2DG model was assumed to be a power law, with the spectral index now fixed to the average value obtained from the fit to the entire data set. The same was done for 47Tuc to allow comparison (and check that no cross-talk occurs between these two neighbouring sources), while the parameters of all other components of our background model were fixed to the average values obtained from the fit to the entire data set. The light curves of the 2DG and 47Tuc models were then derived on a monthly basis. The result is shown in Fig. \ref{fig_light curve}. There is no indication for a flare coming from the direction of the SMC or from 47Tuc \citep[in agreement with the dedicated analysis of the latter, see][]{Abdo:2009f}.\\
\indent Since background blazars obviously would appear as point-like sources, we also derived the light curves for the three point sources of the 3PS model. In this case as well, we found no indication for flaring activity from any of the three directions.

\subsection{Spectrum and flux}
\label{emiss_spec}

\begin{figure}[!t]
\begin{center}
\includegraphics[width=\columnwidth]{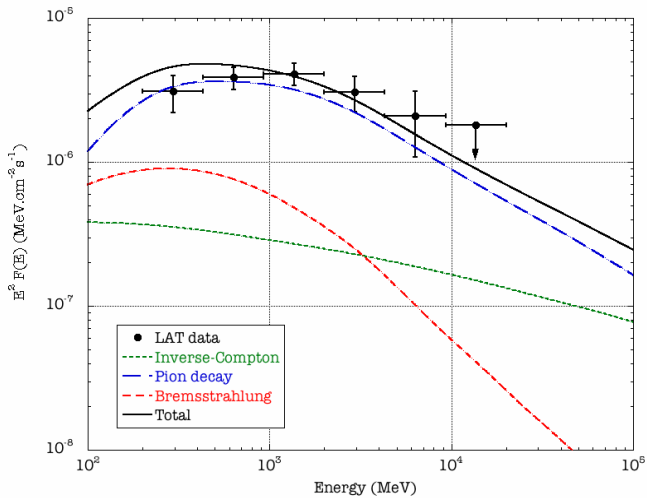}
\caption{Spectrum of the SMC. The LAT data points (in black) come from independent fits of the 2DG plus background models in 6 different energy bins. The curves correspond the components and total of a spectral model of CR-ISM interactions in the SMC fitted to the data using a binned maximum likelihood analysis (see text).}
\label{fig_spec}
\end{center}
\end{figure}

\begin{table*}[!t]
\begin{minipage}[][4.1cm][c]{\textwidth}
\caption{Best-fit spectral parameters obtained from a maximum likelihood analysis with spatial model 2DG}
\begin{center}
\begin{tabular}{|c|c|c|c|c|c|}
\hline
\celltspace Model & Index & Cutoff & F$_{100}$ & F$_{200}$& TS \cellbspace \\
\hline
\celltspace Simple power law & -2.23$^{+0.10}_{-0.10}$$^{+0.07}_{-0.06}$ & - & 4.38$^{+0.71}_{-0.49}$$^{+0.98}_{-0.98}$ & 2.11$^{+0.24}_{-0.24}$$^{+0.12}_{-0.12}$ & 136.6 \cellbspace \\
\hline
\celltspace Exponentially cutoff power law & -1.76$^{+0.22}_{-0.14}$$^{+0.00}_{-0.01}$ & 3.8$^{+3.6}_{-1.3}$$^{+1.8}_{-0.8}$ & 3.66$^{+0.74}_{-0.66}$$^{+0.34}_{-0.60}$ & 1.90$^{+0.26}_{-0.26}$$^{+0.11}_{-0.13}$ & 142.6 \cellbspace \\
\hline
\end{tabular}
\end{center}
\label{tab_spec}
Note to the table: From left to right, table columns are the spectral model, the spectral index, the cutoff energy (in GeV), the photon flux in the 100\,MeV-500\,GeV and 200\,MeV$-$20\,GeV range (both in 10$^{-8}$\,ph\,cm$^{-2}$\,s$^{-1}$), and the TS value. Quoted errors are statistical and systematic respectively.
\end{minipage}
\end{table*}
\indent So far the analysis was done assuming that the spectrum of the SMC emission is well described by a power law. To determine the spectrum of the gamma-ray emission from the SMC without any assumption on the spectral shape, we fitted our background and 2DG models to the data independently in 6 logarithmically-spaced energy bins covering the energy range 200 MeV - 20 GeV. The result is shown in Fig. \ref{fig_spec}. The spectrum is relatively flat ($\Gamma \sim 2$) over most of the energy range, with a possible maximum around one GeV and a break or cutoff around a few GeV.\\
\indent To determine the integrated gamma-ray flux of the spectrum, we fitted both simple power-law and exponentially cutoff power-law spectral models of the form $N(E) = k\,(E/E_0)^{-\Gamma} \exp(-E/E_c)$ to the data by means of a binned maximum likelihood analysis over the energy range 200 MeV - 20 GeV. This analysis is more reliable than fitting the spectra of Fig. \ref{fig_spec} directly since it accounts for the Poissonian statistics of the data. The spectral parameters thus obtained are listed in Table \ref{tab_spec}. The best-fit model is the exponentially cutoff power law. The improvement over the simple power-law model is $\Delta$TS=6 for one more degree of freedom, which corresponds to a significance of 2.4$\sigma$ only. This is most likely because the limited statistics at high energies prevent the clear detection of a cutoff in the spectrum. In the following, we will use the flux value associated with the exponentially cutoff power-law model because it is more relevant to the physical sources we will consider than the simple power-law model: we indeed discuss CRs-ISM interactions as a likely source of the SMC emission and the corresponding radiation is usually dominated by a $\sim$GeV bump due to $\pi^0$ decay, and this feature is better represented with an exponentially cutoff power-law model (see Sect. \ref{smcsrc_cr} and Fig. \ref{fig_spec}); alternatively, we consider pulsars as another plausible source of the SMC emission and the typical spectrum of the latter is characterised by a cutoff at a few GeV (see Sect. \ref{smcsrc_pulsar}). We note that the relatively small difference between both spectral models ensures that the morphological analysis of the SMC emission presented in \ref{morpho}, and performed with an assumed simple power-law spectral shape, remains valid.\\
\indent The integrated flux of the best-fit model is (1.9 $\pm$0.2) $\times$ 10$^{-8}$\,ph\,cm$^{-2}$\,s$^{-1}$ for photons in the 200\,MeV$-$20\,GeV range, and (3.7 $\pm$0.7) $\times$ 10$^{-8}$\,ph\,cm$^{-2}$\,s$^{-1}$ for photons in the 100\,MeV-500\,GeV range, all uncertainties being purely statistical. Since we restricted our analysis to photons $>200$\,MeV the extrapolation down to 100\,MeV may introduce a systematic uncertainty if the low-energy spectrum of the source deviates from our fitted model. In addition, systematic uncertainties on the LAT instrument response function translate into systematic uncertainties on the spectral parameters that are given in Table \ref{tab_spec}. The $>100$\,MeV flux is provided to facilitate comparison with the upper limit from CGRO/EGRET. \citet{Sreekumar:1993} gives an upper limit of 5.0 $\times$ 10$^{-8}$\,ph\,cm$^{-2}$\,s$^{-1}$ from the first years of observations of the SMC with CGRO/EGRET, which is compatible with our extrapolated flux, whatever the spectral model used.

\section{Gamma-ray sources in the SMC}
\label{smcsrc}

\indent Cosmic-ray interactions with the ISM may well be the primary source of gamma rays in the SMC. The resulting emission is expected to be intrinsically diffuse in nature and our analysis indeed suggests that the gamma-ray emission from the SMC does not originate in a small number of individual point sources. When trying to account for the emission from the SMC with a combination of point sources, we detected only three point sources with significance above 2$\sigma$ within the boundaries of the galaxy. In contrast, we found a model with less degrees of freedom that resulted in a larger TS value when we allowed for the source to be extended (model 2DG). The SMC emission is therefore more likely diffuse in nature. Alternatively, it could be composed of a large number of unresolved and faint sources that can not be detected individually by {\em Fermi}/LAT. In the following, we examine these two possibilities in turn.

\subsection{CRs - ISM interactions}
\label{smcsrc_cr}

In this section, we make the assumption that a large fraction of the gamma-ray emission from the SMC indeed originates from CRs interacting with the interstellar gas and radiation field. We first computed the average integrated $>100$\,MeV gamma-ray emissivity per hydrogen atom of the SMC using the following relation:
\begin{linenomath}
\begin{equation}
q_{\gamma}^{> 100\,\textrm{MeV}} = \Phi_{\gamma}^{> 100\,\textrm{MeV}} \, \frac{m_{p}}{M_{gas}} \,d^{2}
\label{eq_emiss_I}
\end{equation}
\end{linenomath}
where $\Phi_{\gamma}$ is the integrated $>100$\,MeV photon flux from the SMC, $M_{gas}$ is the total gas content of the galaxy, $d$ is the distance to the galaxy and $m_{p}$ is the proton mass. In a more practical form (where $>$100\,MeV is implied):
\begin{linenomath}
\begin{align}
q_{\gamma} = 8.0\,10^{-30} \frac{\Phi_{\gamma}}{10^{-7}\,\textrm{ph}\,\textrm{cm}^{-2}\,\textrm{s}^{-1}} \left( \frac{d}{1\,\textrm{kpc}} \right)^{2} \left( \frac{M_{gas}}{10^{8}\,\textrm{M}_{\odot}} \right)^{-1}
\label{eq_emiss_II}
\end{align}
\end{linenomath}
where the emissivity $q_{\gamma}$ is in ph\,s$^{-1}$\,sr$^{-1}$\,H$^{-1}$. With Eq. \ref{eq_emiss_I} and \ref{eq_emiss_II}, we focus on the gamma-rays produced by the interaction of CRs with interstellar matter. These are produced by two processes: decay of neutral pions $\pi^0$ created by inelastic collisions between CR nuclei and ISM gas particles, or Bremsstrahlung of CR electrons and positrons. In the local Galactic conditions, the contribution of Bremsstrahlung becomes small beyond a few 100\,MeV \citep[see for instance][]{Abdo:2009e}, and we will therefore consider in the following that gamma-rays from interactions with interstellar matter are mostly due to the hadronic component of CRs. The questions of the average electron spectrum and electron-to-proton ratio in the SMC are addressed in Sect. \ref{crpop_synch} and \ref{crpop_gam}.\\
\indent The SMC is a gas-rich system \citep[compared to typical spiral galaxies, see][]{Hunter:1997}. The total mass of neutral atomic hydrogen (HI) has been determined from 21\,cm radio observations with the Parkes and ATCA facilities; after correction for self-absorption, it was found to amount to 4.2 $\times 10^{8}$\,M$_{\odot}$ \citep{Stanimirovic:1999}. The total mass of molecular hydrogen (H$_{2}$) has been determined from far-infrared observations with the Spitzer space telescope and found to be 3.2 $\times 10^{7}$\,M$_{\odot}$. The ratio of molecular to atomic gas is therefore 2-3 times lower than in spiral galaxies \citep{Leroy:2007}. We thus obtain a HI+H$_{2}$ gas mass of $\sim 4.5 \times 10^{8}$\,M$_{\odot}$. This estimate, combined with the gamma-ray flux found in \ref{emiss_spec}, translates into an average gamma-ray emissivity $>100$\,MeV of (2.5 $\pm$0.5) $\times$ 10$^{-27}$\,ph\,s$^{-1}$\,sr$^{-1}$\,H$^{-1}$ for an adopted distance to the SMC of 61.5\,kpc \citep{Cioni:2000,Hilditch:2005,Keller:2006}, with the uncertainty reflecting only the uncertainty on the gamma-ray flux.\\
\indent The average gamma-ray emissivity of the SMC is a factor of 6-7 times smaller than the emissivity $q_{\gamma} = (1.63 \pm 0.05) \times 10^{-26}$ ph\,s$^{-1}$\,sr$^{-1}$\,H$^{-1}$ that has been determined with {\em Fermi}/LAT for the local ISM of our own Galaxy \citep{Abdo:2009e}. However, these quantities are not directly comparable for the reasons we list below.\\
\indent First, our derivation of the average gamma-ray emissivity per H-atom of the SMC implicitly assumes that all gamma-rays are produced by $\pi^0$ decay after hadronic interactions. Yet, we know from the study of the Galactic high-energy diffuse emission that another process of leptonic origin, the inverse-Compton scattering, can account for a certain part of the flux in the 200 MeV - 20 GeV energy range. The contribution of inverse-Compton was subtracted prior to the determination of the local emissivity using predictions from the GALPROP model, but the average emissivity of the SMC was computed from a mix of $\pi^0$ decay and inverse-Compton gamma-rays in unknown proportions.\\
\indent Second, some of the observed emission from the SMC may arise from unresolved point sources like pulsars, and this would overestimate the flux due to CRs interacting with interstellar matter (see Sect. \ref{smcsrc_pulsar}). In contrast, the local emissivity was determined for high-latitude regions of the Galaxy, mostly within 1\,kpc from the solar system. The observations were cleaned for point sources (at least for the point sources detected by the LAT over the first 6 months) and it seems reasonable to assume that the emission from the small volume under investigation was left relatively free of contamination by isolated objects.\\
\indent Last, the emissivity is computed from the mass of hydrogen and as evident from Eq. \ref{eq_emiss_I}, the higher the mass for a given gamma-ray flux, the lower the emissivity. The value we used for the SMC (4.5\,10$^{8}$\,M$_{\odot}$) is probably a lower limit of the total mass of gas in the galaxy. First, we neglected the ionised gas, which can account for up to 10\% \citep[in the case of the MW, see][]{Cordes:2002}. More important is the possibility that the total SMC gas mass is not completely traced by the usual methods. This additional gas could be in the form of cold atomic gas being optically thick (thus not detectable at 21\,cm), or molecular gas free of CO (thus not detectable at 2.6\,mm). The latter scenario would be particularly relevant for a low-metallicity system such as the SMC, in which CO may be underabundant and/or not efficiently shielded from dissociating radiation because of a low dust content. In contrast, the high-latitude observations of the solar neighbourhood are probably not too much affected by this dark gas because the latter surrounds the CO clouds and is thus mostly confined to the Galactic plane. In their analysis of a region of the second Galactic quadrant, \citet{Abdo:2009h} found that most of the dark gas is confined to Galactic latitudes $|b| \leq 20\deg$, whereas \citet{Abdo:2009e} derived the local emissivity from a region of the third Galactic quadrant with $|b| \geq 22\deg$. In addition, it is worth noting that \citet{Abdo:2009e} found good agreement between the emissivity inferred from the diffuse gamma-ray emission and the emissivity computed from the measured spectrum of CRs without resorting to any dark gas contribution.\\
\indent On the whole, this means that the true average emissivity for the SMC is very likely below the above-mentioned value. The average gamma-ray emissivity of the SMC should therefore be regarded as being \textit{at least} 6-7 times smaller than the local value. We emphasise, however, that depending on the actual relative distributions of gas and CRs in the SMC, the emissivity may reach higher values in some places. This was clearly illustrated in the recent {\em Fermi}/LAT study of the LMC. The average emissivity of the LMC is found to be 2-4 times smaller than the local value. Yet, the gamma-ray emission from the LMC is observed to be poorly correlated with the gas density, and this translates into a gamma-ray emissivity being higher than the local value in several places \citep[see Fig. 10 in][]{Abdo:2009g}. For the SMC, the current photon statistics are insufficient to build an emissivity map of the galaxy, so we must confine ourselves to the global average.\\
\indent In comparing the gamma-ray emissivities per H-atom of the local Galaxy and the SMC, one should also not forget the effect of the chemical composition of both the ISM and the CRs. The latter is usually taken into account through the so-called \textit{nuclear enhancement factor} $\epsilon_{N}$, which is a multiplicative factor applied to the emissivity due to proton-proton reactions only. Recent calculations by \citet{Mori:2009} yielded $\epsilon_{N} = 1.85$ for CRs having the composition measured at Earth and interacting with an ISM of solar metallicity. The Z= 0.2 Z$_{\odot}$ metallicity of the SMC \citep{Dufour:1984} corresponds to $\epsilon_{N}=1.72$, so we can consider that the effect of the low-metallicity of the SMC on the gamma-ray emissivity is negligible.\\
\indent Our result, together with the above arguments, indicates that the average density of CR nuclei in the SMC is at least 6-7 times smaller than in the MW at the position of the solar system if the energy spectrum of both populations of CR nuclei are assumed similar. Yet, there is no reason for the CR spectrum not to vary from one galaxy to another. Indeed, the spectrum of the large-scale, steady-state population of CRs is very likely shaped primarily by transport, since we do not expect the diffusive shock acceleration mechanism to be strongly different, and transport of CRs may be highly dependent on characteristics of the host galaxy like geometry, magnetic field strength and topology, or level of interstellar and magnetic field turbulence. The actual spectrum of CR nuclei can be derived from the shape of the gamma-ray spectrum. However, our data do not have sufficient statistics to be conclusively constraining in this regard. These issues are addressed in more detail in \ref{crpop_gam} through modelling of the gamma-ray emission from CR-ISM interactions in the SMC.

\subsection{Isolated sources}
\label{smcsrc_pulsar}

\indent The current {\em Fermi}/LAT observations of the SMC are limited in their capability to rule out the possibility that a sizeable fraction of the diffuse gamma-ray emission is due to a large number of unresolved and faint sources. In this section, we estimate that contribution to the observed emission.\\
\indent The 1FGL catalog contains a total of 1451 sources, about half of which were associated with candidate sources. Most of the associated sources are active galactic nuclei, and the remaining associated sources are mostly Galactic objects \citep[see for instance][]{Abdo:2009b}. Although there is still a large number of unidentified sources, many of which lie in the Galactic plane, the Galactic population of gamma-ray sources seems to be dominated by pulsars. In the following, we will therefore focus on this class of object as possible sources of gamma-ray emission in the SMC.\\
\indent The first {\em Fermi}/LAT catalogue of gamma-ray pulsars based on the first six months of LAT observations contains 46 objects, most of which are normal non-recycled pulsars \citep{Abdo:2009i}. From distance estimates and using a flux correction factor $f_{\Omega}=1$, the observed fluxes were converted into gamma-ray luminosities $>$ 100\,MeV approximately ranging from 10$^{34}$ to 10$^{36}$\,erg\,s$^{-1}$ for most normal non-recycled pulsars. The inferred luminosities suffer from large uncertainties on the distance estimates, which are manifest from the $> 100\%$ efficiencies implied for several objects. Another source of error is the assumption of a uniform beaming factor $f_{\Omega}=1$ across the sky, which is not realised in many emission models and thus overestimates the gamma-ray luminosity in some cases. The characteristic ages for most normal non-recycled pulsars range from 10$^4$ to 10$^6$\,yrs, and their typical spectrum has been found to be an exponentially cut-off power law with index 1.4 and cut-off energy 2.2\,GeV. In the following, we will assume that normal gamma-ray pulsars have a mean luminosity of 10$^{35}$\,erg\,s$^{-1}$ and a mean lifetime of 10$^5$\,yrs. Adopting the above mentioned mean spectrum, the energy luminosity of 10$^{35}$\,erg\,s$^{-1}$ translates approximately into a photon luminosity of 10$^{38}$\,ph\,s$^{-1}$.\\
\indent \citet{Crawford:2001} estimated the number of active radio pulsars in the SMC to 5100 $\pm$3600, under the assumption that radio pulsars form at the same rate as SNII and have a mean lifetime of 10\,Myr. This would imply 51 $\pm$36 gamma-ray pulsars, under the hypothesis that all young pulsars are gamma-ray emitters with a lifetime of 0.1\,Myr. We currently know 6 pulsars in the SMC\footnote{These 6 objects actually are those listed in the ATNF catalogue referred to previously, but there are in fact many more known pulsars in the SMC. \citet{Galache:2008} lists about 30 X-ray pulsars that are presumably parts of Be/X-ray binaries. Yet, these pulsars are not expected to be powerful gamma-ray emitters because of their long periods $\geq$10s.}, 5 discovered in radio and 1 in X-rays \citep{Manchester:2005,Lamb:2002}. Most of these pulsars have characteristic ages above 1\,Myr and may therefore be too old for substantial gamma-ray emission. A possible exception is J0100-7211, which is an Anomalous X-ray Pulsar candidate with an estimated age of 6800\,yrs \citep{McGarry:2005}.\\
\indent A mean luminosity of 10$^{38}$\,ph\,s$^{-1}$ per pulsar in the SMC translates into a photon flux at Earth of $\sim$ 2\,10$^{-10}$\,ph\,cm$^{-2}$\,s$^{-1}$. The 51 $\pm$36 gamma-ray pulsars would therefore give rise to a gamma-ray flux of $(1.0 \pm0.7)\,10^{-8}$\,ph\,cm$^{-2}$\,s$^{-1}$, to be compared with the observed value of $(3.7 \pm0.7)\,10^{-8}$\,ph\,cm$^{-2}$\,s$^{-1}$. Taking a lifetime of 1\,Myr instead of 0.1\,Myr for gamma-ray pulsars leads to the conclusion that the entire SMC emission can be accounted for by pulsars. Yet, these estimates suffers from large uncertainties, first in the birth rate of gamma-ray pulsars and then for the gamma-ray emission from a typical pulsar population. A more reliable calculation would require a complete population synthesis including, in particular, the time evolution of the gamma-ray luminosity. However, such a simulation is quite sensitive to the as yet unsettled magnetospheric processes governing high-energy emission \citep[see][and references therein]{Abdo:2009i}. Nevertheless, the above calculation shows that a substantial fraction of the observed emission could be accounted for by normal pulsars. Adding other objects like millisecond pulsars would further increase the contribution of isolated sources.\\
\indent On the whole, this means that the density of CRs in the SMC may be even smaller than the values discussed previously, and hence significantly below the local value. This also implies that the gamma-ray emission of the SMC may be of a different composition compared to the Milky Way. In the Galaxy, only 10-20\% of the gamma-ray emission can be attributed to isolated sources while the rest is of diffuse nature, but in the SMC the proportions may be inverted. Such a scenario would require a lower mean SN rate per unit volume (to form gamma-ray pulsars without increasing the mean CR injection rate per unit volume) and/or a lower confinement of CRs (to reduce the accumulation of particles in the ISM). We will discuss this in more detail in \ref{crpop_trans}.\\
\indent On that issue, we note that the refinement of the emission morphology that will result from increased exposure may allow the assessment of the relative contributions of each kind of source. Also, the typical spectrum of pulsars has a specific shape that may not be consistent with the better-defined high-energy spectrum that will follow from better statistics.

\section{The CR population of the SMC}
\label{crpop}

In this section, we focus on the implications in terms of CR population of the detection of the SMC in gamma-rays. We again assume that the entire emission is due to CR-ISM interactions and we derive the characteristics of the CR population of the SMC using a more detailed approach than the simple emissivity calculation performed in Sect. \ref{smcsrc_cr}. In order to broaden the issue, we investigated the population of CR electrons in the SMC. Compared to CR nuclei, CR electrons undergo severe radiative energy losses. While the main loss mechanism for nuclei above 1\,GeV is escape from the host galaxy, the CR electrons lose the bulk of their energy through inverse-Compton scattering of the interstellar radiation field and synchrotron radiation on the galactic magnetic field. Hence, the most energetic CR electrons do not travel far from their source regions before losing most of their energy. By considering the conditions for the CR nuclei and electrons in the SMC, it may be possible to assess if the lower density of CRs in the SMC is due to injection or to transport. In addition, we discuss the aspects of injection and/or transport that may drive the large-scale properties of the CR population of the SMC and their differences with respect to the Galaxy.

\subsection{Constraints from synchrotron emission}
\label{crpop_synch}

\indent When accelerated in the large-scale magnetic field of a galaxy, CR electrons manifest themselves through diffuse synchrotron emission. From \citet{Longair:1994}, the synchrotron radiation spectrum of a power-law energy distribution of electrons is (with some changes in the units):
\begin{linenomath}
\begin{equation}
\label{eq_synchro_emiss}
J(\nu) = 2.344 \times 10^{-29} \, f(p) \, B^{\frac{p+1}{2}} \, K_{e} \, \left( \frac{3.217 \times 10^{7}}{\nu} \right)^{\frac{p-1}{2}}
\end{equation}
\end{linenomath}
where $\nu$ is the radiation frequency in Hz, $J(\nu)$ is the emissivity per unit volume in W\,m$^{-3}$\,Hz$^{-1}$, $B$ is the magnetic field in $\mu$G, $N_{e}(E)=K_{e}(E/E_0)^{-p}$ is the energy spectrum of CR electrons in cm$^{-3}$\,GeV$^{-1}$ and $f(p)$ is a numerical factor depending on the slope of the particle spectrum. Then, the intensity of a given object is given by:
\begin{linenomath}
\begin{equation}
\label{eq_synchro_flux}
F(\nu) = J(\nu) \, \frac{d\Omega}{4 \pi} \, dl
\end{equation}
\end{linenomath}
where $d\Omega$ is the solid angle subtended by the diffuse source, in sr, and $dl$ is the mean depth of the emitting medium along the line of sight, in m. The intensity $F(\nu)$ is in W\,m$^{-2}$\,Hz$^{-1}$ (which corresponds to 10$^{26}$\,Jy).\\
\indent The radio continuum emission from galaxies usually has a steep spectrum at lower frequencies, due to synchrotron emission from non-thermal particles, followed by a flatter spectrum at higher frequencies, due to free-free emission from thermal particles (with a typical index of -0.1). \citet{Haynes:1991} determined a spectral index of -0.85 $\pm$0.10 for the non-thermal radio continuum emission of the SMC, with a flux density\footnote{The total flux density at 1.4\,GHz is 42 $\pm$6\,Jy and we subtracted 20\% for the contribution of thermal emission (as estimated by the authors at 1\,GHz).} of 34 $\pm$6\,Jy at 1.4\,GHz.\\
\indent From Eq. \ref{eq_synchro_emiss}, the radio measurements give \mbox{$p$= -2.7 $\pm$0.2} for the average spectral index of the CR electron distribution of the SMC. The normalisation of the CR electron spectrum can be obtained from the flux density at a given frequency if we know the magnetic field strength and the spatial distribution of the synchrotron-emitting medium (see Eq. \ref{eq_synchro_emiss}).\\
\indent The large-scale magnetic field of the SMC was recently studied by \citet{Mao:2008} using optical starlight polarisation and rotation measures for background extragalactic radio sources. The authors determined that the coherent magnetic field in the SMC has a strength of $\sim$ 1.7\,$\mu$G and lies almost entirely in the plane of the sky. The SMC also seems to have a random magnetic field with an estimated strength of $\sim$ 2-3\,$\mu$G depending on the method used. The observed synchrotron emission from the SMC actually arises from the total magnetic field projected onto the plane of the sky $B_{total,\perp}$, and \citet{Mao:2008} estimated this component to have a strength of $\sim$ 3\,$\mu$G.\\
\indent The 3D structure of the SMC is quite controversial and this directly affects the interpretation of the radio synchrotron emission. The size and appearance of the SMC in the sky is biased by observations of young stars and gas, which suggest a very irregular and asymmetric shape while the old stellar population seems to follow a very regular distribution \citep{Gonidakis:2009}. Yet, for the purpose of interpreting radio synchrotron emission from CR electrons, young, massive stars are more relevant than old stars since they are putative sources of CRs. From a visual inspection of the radio maps of the SMC, it appears that the emitting regions can be enclosed in a circle of $5\deg$ in diameter corresponding to a solid angle of $d\Omega=$6\,10$^{-3}$\,sr. However, the majority of the emission likely comes from half of this solid angle. The issue of the extension of the SMC along the line of sight is important as well. Past estimates of the depth of the galaxy range from a few kpc to a few 10\,kpc \citep[see the review by][section 4.2]{Stanimirovic:2004}. Currently, lower values are favoured. The deviations of pulsating red giants in the SMC from the mean period-luminosity relations indicate a distribution over a distance range of 3.2 $\pm$1.6\,kpc \citep{Lah:2005}. This result is reminiscent of the distance dispersion of $\pm$ 3.3\,kpc found by \citet{Welch:1987} from an analysis of Cepheid variables in the SMC (and put forward as an argument that the SMC is not extended beyond its tidal radius and is therefore not undergoing disintegration). In the following, we will adopt a mean depth along the line of sight of $dl=$4\,kpc. In our further calculations, however, we will also consider larger depth such as the 10 $\pm$6\,kpc found by \citet{Mao:2008} when deriving the line of sight magnetic field of the SMC from rotation measures.\\
\indent From the above data and Eqs \ref{eq_synchro_emiss} and \ref{eq_synchro_flux}, we computed the normalisation factor for the CR electron population of the SMC and obtained $K_{e}=2.5\,10^{-12}$\,cm$^{-3}$\,GeV$^{-1}$. This corresponds to an energy density of $w_{CR,e}=3.6\,10^{-3}$\,eV\,cm$^{-3}$ for particles above 1\,GeV with a spectral index of $p$= -2.7. For comparison, the local energy density of CR electrons estimated from the recent {\em Fermi}/LAT measurement is $w_{CR,e} \simeq 6.0\,10^{-3}$\,eV\,cm$^{-3}$ for particles above 1\,GeV with a spectral index of $p$= -3.0 \citep[][assuming the spectral slope measured over the 20\,GeV--1\,TeV range extends down to 1\,GeV]{Abdo:2009a}.\\
\indent The average density of CR electrons in the SMC from this argument is $\sim 60$\% of the Galactic local value. Yet, when taking into account the uncertainties on the radio data (spectral index and intensity), we find that the CR density derived for the SMC undergo relative variations from \mbox{-50\%} to \mbox{+70\%}. Also, the depth of the galaxy may be larger than the 4\,kpc we adopted and assuming a depth of 10\,kpc instead of 4\,kpc would decrease the CR electron energy density by 60\%. On the other hand, the size of the synchrotron emitting region may be smaller by 50\% compared to the value used so far and this would nearly compensate an increased depth.\\
\indent Large uncertainties therefore affect the derivation of the CR electron density in the SMC from radio synchrotron observations. At present, it is consistent with the local value and with a 6-7 reduction factor as observed for CR nuclei (the latter case implying the same electron-to-proton ratio as measured locally). CR electrons also manifest themselves through inverse-Compton gamma-ray emission, which offers an alternative to address the above questions. However, as discussed in the next section, the statistical limitations of the current high-energy observations are not very constraining either.

\subsection{Constraints from gamma-ray emission}
\label{crpop_gam}

\indent We simulated a gamma-ray spectrum for the SMC using a one-zone model of CR-ISM interactions that takes into account $\pi^0$ decay following proton-proton interactions, Bremsstrahlung from CR electrons and inverse-Compton scattering of cosmic-ray electrons on optical, infrared and cosmic microwave background photons. In this model, the various quantities involved in the CR-ISM interactions are represented as average values over a given volume or gas mass, depending of the emission process. For the CR population, we used the proton, electron and positron spectra presented in \citet{Abdo:2009e} for the local Galactic environment, with a global scaling factor to account for the CR density difference between the SMC and the local MW (see below).\\
\indent We calculated the $\pi^0$ production by proton-proton interactions following the prescription of \citet{Kamae:2006}. The $\pi^0$ and Bremsstrahlung emissivities were calculated assuming an average SMC metallicity of $Z=0.2$\,Z$_{\odot}$ \citep{Dufour:1984}, which implies a nuclear enhancement factor of $\epsilon_{N}=1.72$ for the $\pi^0$ emissivity \citep{Mori:2009}. These average emissivities are then multiplied by the the total hydrogen mass of the SMC ($4.5 \times 10^8$\,M$_{\odot}$, see Sect. \ref{smcsrc_cr}) to give the total $\pi^0$ and Bremsstrahlung gamma-ray luminosity.\\
\indent The inverse Compton component was calculated from the method described by \citet{Blumenthal:1970} using the CMB, optical and infrared interstellar radiation fields (ISRFs). The optical and infrared ISRFs were taken from the GALPROP model of the MW \citep{Porter:2008} and averaged over a cylindrical volume with a 15\,kpc radius and a 4\,kpc thickness. The resulting average optical and infrared ISRFs were rescaled according to the stellar luminosity density and observed infrared emission for the SMC \citep[factors of $\sim$1.2 and $\sim$1.25 respectively, see][]{Bot:2004}. The inverse Compton gamma-ray luminosity was then calculated assuming that the CR electrons of SMC interact with the target photons in a cylinder with a 1.6\,kpc radius (radius of the SMC gamma-ray emission, see \ref{morpho_geom}) and a 4\,kpc height (depth of the SMC along the line of sight, see \ref{crpop_synch}).\\
\indent The spectral profile thus obtained is associated with the 2DG intensity distribution to form a complete model for the SMC emission. The scaling factor $r_c$ of our model with respect to the data, which we refer to as the cosmic-ray enhancement factor, is then a direct measure of the average density of CRs in the SMC with respect to that in the vicinity of the Earth. From a binned maximum likelihood analysis, we found $r_c= 0.14$ with a statistical uncertainty of $\pm0.01$. Systematic errors due to uncertainties in the effective area of the instrument amount to $\pm 0.01$, and an additional uncertainty of $\pm$0.01 comes from the uncertainty on the extent of the source (the $\sigma$ of the 2D Gaussian model). The $r_c$ factor obtained from this spectral modelling is consistent with the estimate made in \ref{smcsrc_cr}.\\
\indent From this one-zone model of CR-ISM interactions in the SMC, the assumption of CR spectra and CR electron-to-proton ratio identical to the locally observed values does not conflict with the gamma-ray spectrum.

\subsection{Injection and transport}
\label{crpop_trans}

\indent The average density of CR nuclei in the SMC seems to be at least 6-7 times smaller than in our local environment, if we assume a similar spectrum for the particles. The energy distribution of CR nuclei is poorly constrained by our gamma-ray spectrum. The density of CR electrons in the SMC was found to be $\sim$ 2 times smaller than in our local environment, but due to uncertainties on the radio data and other parameters, it could be down to 10 times smaller or consistent with the local value. The specifics of CR injection and transport in the SMC with respect to the Milky Way lead to lower average CR densities and possibly also to a higher electron-to-proton ratio, although a similar electron-to-proton ratio cannot be excluded.\\
\indent Similar results were obtained in a recent study of the LMC based on {\em Fermi}/LAT observations. We report here the main conclusions of that work for comparison. The gamma-ray emission from the LMC is clearly of diffuse nature, with an extension of angular size $\sim3.5\deg$. The total flux from the LMC is \mbox{(2.6 $\pm$0.2 $\pm$0.4) $\times$ 10$^{-7}$\,ph\,cm$^{-2}$\,s$^{-1}$}, which is about 7 times the flux we obtained for the SMC. The spectrum was found to be well represented by a power law with an index of about -2.0 and a cut-off at a few GeV, quite similar to what we obtained for the SMC. These observations, combined with the properties of the LMC, imply a CR density that is 2-4 times smaller than the local Galactic value, possibly even less. Regarding CR electrons in the LMC, the authors did not explicitly estimate their density but they showed that the assumption of a CR electron-to-proton ratio identical to the locally observed value does not conflict with the gamma-ray spectrum or the radio synchrotron observations of the LMC.\\
\indent A lower density of CR nuclei in a steady-state system can result from a lower rate of CR injection and/or a reduced confinement in the volume of the galaxy. To first order, the injection rate can be simply related to the rate of supernovae or to the rate of star formation \citep{Pavlidou:2001}, because we do not expect the diffusive shock acceleration process occurring in supernova explosions to strongly differ from one galaxy to another. Conversely, CR confinement may be highly dependent on some characteristics of the galaxy like geometry, magnetic field strength and topology or the presence of a galactic wind and/or halo. In the following, we discuss those specifics of the SMC that may explain the lower CR densities.\\
\indent The star formation history derived by \citet{Harris:2004} from \textit{UBVI} photometry of over 6 million SMC stars show that the star formation rate (SFR) in the SMC over the last Gyr has an underlying constant value of $\sim$0.1\,M$_{\odot}$\,yr$^{-1}$, with episodes of enhanced activity at 60 and 400\,Myr where the value was increased by factors of 2-3. Based on far-infrared dust emission and H$\alpha$ emission, \citet{Wilke:2004} deduced a present-day SFR of $\sim$0.05\,M$_{\odot}$\,yr$^{-1}$, in good agreement with the value of \citet{Harris:2004} and with previous estimates. By comparison, a review of the SFR estimates for the Milky Way by \citet{Diehl:2006} shows that a value of 4 $\pm$2\,M$_{\odot}$\,yr$^{-1}$ can be adopted. The SFR in our Galaxy is therefore about 40 times higher than in the SMC. We note that a similar scaling is found for the supernova rates: the commonly adopted value for the Milky Way is $\sim$3 per century, while the estimates for the SMC are $\sim$0.1 per century \citep{Tammann:1994}.\\
\indent Yet, for the issue of steady-state CR density in a given system, it is not the absolute rate of star formation or supernovae that matters, but the rate by unit volume. Estimating the volumes in which star formation occurs is not an easy task, especially for the SMC, the exact morphology of which is still unclear (see Sect. \ref{crpop_synch}). So instead of trying to compute these volumes, we used the interstellar radiation field density as an indicator of the star formation per unit volume. The average temperature of the dust is observed to be higher in the SMC than in the Milky Way, suggesting a stronger interstellar radiation field \citep{Stanimirovic:2000,Wilke:2004}. This could be due to the lower metallicity of the SMC, but may also be the consequence of a higher SFR per unit volume. If the rate of CR injection per unit volume actually is higher in the SMC than in the Galaxy, then the average lower CR densities we inferred for the SMC could be due to some properties of their transport in the ISM.\\
\indent In a basic picture, the diffusion of CRs in the ISM is governed by two parameters: the spatial diffusion coefficient $D_{xx}$ and the confinement volume characteristic size $l_{conf}$ (both parameters being energy-dependent). The main ISM characteristic thought to affect $D_{xx}$ is the $B / \delta B$ parameter \citep[see the review by][]{Strong:2007}. From the estimate of the random and ordered magnetic field strengths in the SMC by \citet{Mao:2008}, there is no reason to think that $D_{xx}$ in the SMC could be substantially different from $D_{xx}$ in the Galaxy. We are therefore left with the confinement volume size $l_{conf}$ as the potential explanation for the lower CR density of the SMC.  Below, we make a first-order comparison of this parameter for the SMC and MW.\\
\indent Neglecting energy losses, the steady-state average CR density in a galaxy can be approximated by \citep[see for instance][]{Pavlidou:2001}:
\begin{linenomath}
\begin{equation}
N(E)=q(E) \, \tau_{esc}(E)
\end{equation}
\end{linenomath}
where $\tau_{esc}$ is the characteristic escape time and $q(E)$ the source term. For a diffusive process, the confinement volume size can be approximated by:
\begin{linenomath}
\begin{equation}
l_{conf} = \sqrt{D_{xx} \, \tau_{esc}}
\end{equation}
\end{linenomath}
From spectral modelling, we found that the mean CR density in the SMC is $\sim$15\% or less the CR density in the local Galaxy. Assuming a similar diffusion coefficient and a potentially higher injection rate in the SMC, the above equations imply a mean CR confinement time in the SMC of $\sim$15\% the Galactic value, which is typically $\sim 10^{7}$\,yrs, and a mean CR confinement volume size in the SMC at least 3 times smaller than in the MW.\\
\indent Similar conclusions can probably be drawn for the LMC. Discussing the possible physical reasons to this is beyond the scope of that paper. It may be a simple size effect like the surface-to-volume ratio (injection being proportional to volume while escape is proportional to surface), or a more sophisticated process involving the equilibrium of the various galactic components (CRs, gas and magnetic field resisting to gravitational attraction). It should be noted that the comparison of the Magellanic Clouds to the Milky Way in terms of a large-scale characteristic such as the CR population may be complicated by the fact that smaller galaxies are probably more affected by the stochasticity of star formation and by tidal interactions with their neighbours. The concept of steady-state CR population may simply not apply in these cases.\\
\indent These rough estimates, however, hold for a mean CR density computed under the assumption of a uniformly distributed CR population. As mentioned in \ref{smcsrc_cr}, if CRs are poorly correlated with the gas, then the CR density can be expected to vary over the SMC volume and may well reach values above the local Galactic one. If some correlation of the CRs with supergiant shells exists, as mentioned in \ref{morpho_correl} and as was found to be the case in the LMC \citep{Abdo:2009g}, the mean CR confinement time and volume may well differ from the average values computed above. For instance, the dynamical ages of the three supergiant shells of the SMC are $\sim1.5 10^{7}$\,yrs \citep{Stanimirovic:1999}. If CRs can be efficiently confined in the superbubbles interiors for this duration, the CR confinement time would be about an order of magnitude above the average value computed above. Such a scenario may be further explored observationally by looking for a correlation of the GeV signal with diffuse X-ray emission, the latter having the potential to reveal the hot gas filling the superbubble cavities. On the theoretical side, additional works are needed to better understand the transport of CRs in a hot, tenuous and presumably highly turbulent medium and determine the time scale over which CRs can be retained within supergiant shells.

\section{Conclusion}
\label{conclu}

\indent From the first 17 months of {\em Fermi}/LAT observations, we obtained the first detection of the SMC in high energy gamma-rays. The flux in the 200\,MeV$-$20\,GeV energy range of the data is (1.9 $\pm$0.2) $\times$ 10$^{-8}$\,ph\,cm$^{-2}$\,s$^{-1}$, and increases to (3.7 $\pm$1.7) $\times$ 10$^{-8}$\,ph\,cm$^{-2}$\,s$^{-1}$ when extrapolated to the 100\,MeV-500\,GeV range from our best-fit spectral model. The latter value is consistent with the upper limit derived from the CGRO/EGRET data.\\
\indent The emission is steady and from an extended source $\sim3\deg$ in size. It is not clearly correlated with the distribution of massive stars or neutral gas, nor with known pulsars or supernova remnants. However, a certain correlation of the gamma-ray emission with supergiant shells is observed. In the recent {\em Fermi}/LAT study of the LMC, the gamma-ray emission was observed to be poorly correlated with gas density but strongly correlated with tracers of massive star forming regions such as ionised gas, Wolf-Rayet stars, and supergiant shells. Better statistics from increased exposure may reveal patterns in the SMC emission and clarify the differences and similarities in the gamma-ray emission of the two Magellanic Clouds.\\
\indent If the gamma-rays from the SMC are produced by CR-ISM interactions, the observed flux implies an average density of CR nuclei in the SMC of $\sim$15\% the value measured locally in the MW. This value should be considered an upper limit since a substantial fraction of the emission may be due to the population of high-energy pulsars of the SMC and the gas mass used for the calculation may be underestimated. The average density of CR electrons derived from radio synchrotron observations of the SMC is consistent with the same reduction factor, but large uncertainties on the radio data and on the geometry of the galaxy also allow higher values. From our current knowledge on the SMC, such a low CR density does not seem to be due to a lower rate of CR injection but may indicate some dependence on CR transport effects, such as a smaller CR confinement volume characteristic size.

\begin{acknowledgement}
The \textit{Fermi} LAT Collaboration acknowledges generous ongoing support from a number of agencies and institutes that have supported both the development and the operation of the LAT as well as scientific data analysis. These include the National Aeronautics and Space Administration and the Department of Energy in the United States, the Commissariat \`a l'Energie Atomique and the Centre National de la Recherche Scientifique / Institut National de Physique Nucl\'eaire et de Physique des Particules in France, the Agenzia Spaziale Italiana and the Istituto Nazionale di Fisica Nucleare in Italy, the Ministry of Education, Culture, Sports, Science and Technology (MEXT), High Energy Accelerator Research Organization (KEK) and Japan Aerospace Exploration Agency (JAXA) in Japan, and the K.~A.~Wallenberg Foundation, the Swedish Research Council and the Swedish National Space Board in Sweden. Additional support for science analysis during the operations phase is gratefully acknowledged from the Istituto Nazionale di Astrofisica in Italy and the and the Centre National d'\'Etudes Spatiales in France.
\end{acknowledgement}

\bibliographystyle{aa}
\bibliography{/Users/pierrickmartin/Documents/MyPapers/biblio/Pulsars,/Users/pierrickmartin/Documents/MyPapers/biblio/SMC,/Users/pierrickmartin/Documents/MyPapers/biblio/CosmicRaySources,/Users/pierrickmartin/Documents/MyPapers/biblio/CosmicRayTransport,/Users/pierrickmartin/Documents/MyPapers/biblio/GalaxyObservations,/Users/pierrickmartin/Documents/MyPapers/biblio/DataAnalysis,/Users/pierrickmartin/Documents/MyPapers/biblio/Fermi,/Users/pierrickmartin/Documents/MyPapers/biblio/Books,/Users/pierrickmartin/Documents/MyPapers/biblio/SNobservations,/Users/pierrickmartin/Documents/MyPapers/biblio/26Al&60Fe}

\end{document}